\documentclass[intlimits,twoside,a4paper]{article}

\usepackage{amsmath,amssymb}
\usepackage{graphicx}

\usepackage[T2A]{fontenc}
\usepackage[cp1251]{inputenc}

\usepackage[eqsecnum]{cmpj2}

\issue{2015}{18}{4}{43601}
\doinumber{10.5488/CMP.18.43601}

\title[Elastic and thermodynamic properties of Pt$_{3}$Al with the L1$_{2}$ structure under high pressure]%
{Theoretical study of the elastic and thermodynamic properties of Pt$_{3}$Al with the L1$_{2}$ structure \\ under high pressure}
\author[N. Wei, Ch. Zhang, S. Hou]{N. Wei\refaddr{label1},
        Ch. Zhang\refaddr{label2}, S. Hou\refaddr{label2}}
\addresses{
\addr{label1} Anhui Sanlian University, Hefei 230601, China
\addr{label2} Key Laboratory of Materials Physics, Institute of Solid State Physics, Chinese Academy
of Sciences, \\ Hefei 230031, China
}

\date{Received December 18, 2014, in final form July 10, 2015}
\authorcopyright{N. Wei, Ch. Zhang, S. Hou, 2015}

\begin{document}
\maketitle
\begin{abstract}
In this work, the elastic and thermodynamic properties of Pt$_{3}$Al under high pressure are investigated using density functional theory within the generalized gradient approximation. The results of bulk modulus and elastic constants at zero pressure are in good agreement with the available theoretical and experimental values. Under high pressure, all the elastic constants meet the corresponding mechanical stability criteria, meaning that Pt$_{3}$Al possesses mechanical stability. In addition, the elastic constants and elastic modulus increase linearly with the applied pressure. According to the Poisson's ratio $\nu$ and elastic modulus ratio ($B/G$), Pt$_{3}$Al alloy is found to be ductile, and higher pressure can significantly enhance the ductility. Those indicate that the elastic properties of Pt$_{3}$Al will be improved under high pressure. Through the quasi-harmonic Debye model, we first successfully report the variations of the Debye temperature $\Theta_\textrm{D}$, specific heats $C_{P}$, thermal expansion coefficient $\alpha$, and Gr\"{u}neisen parameter $\gamma$ under pressure range from 0 to 100~GPa and temperature range from 0 to 1000~K.
\keywords first-principles, elastic properties, thermodynamic properties
\pacs 61.82.Bg, 62.20.dc, 71.20.Be, 71.15.Mb
\end{abstract}

\section{Introduction}

 The high-temperature and ultra-high temperature structural materials, platinum group metal alloys (PGMS) are of great interest due to the high melting point, high strength and exceptional environmental resistance \cite{Jiang, Yamabe, Meschel, Wang}. They have been applied to many important industrial fields, such as catalysts, high temperature structural materials, special solder and shape memory alloys. In recent years, they have been receiving increasing concerns from researchers and have been extensively investigated by experiments and theoretical calculations. Wu et al. have studied the phase diagram of Al-Pt system using the CALPHAD method \cite{Wu}. Then, J. Feng et al. employed the density functional theory (DFT) method  to investigate the stability, thermal and mechanical properties of Pt-Al intermetallic compounds \cite{Feng}. They found that their Poisson's ratio varies from 0.26 to 0.39 and the bonds in the compounds are mainly of metallic and covalent type, which is the same as in Zr-Al alloys \cite{Duan}. As one of the most valuable intermetallic compounds, the L1$_{2}$ phase has been also extensively investigated. Chauke et al. have performed DFT calculations to examine the heats of formation, elastic modulus and the phonon dispersion curves of four different structure-types of Pt$_{3}$Al at absolute zero pressure \cite{Chauke}. The tetragonal DO$_\textrm{c}$ structure is found to collapse to the cubic L1$_{2}$ structure. Norihiko et al. have discussed the single crystals of Pt$_{3}$Al with the L1$_{2}$ structure from 77 to 1073~K \cite{Norihiko}. Gornostyrev et al. have employed first-principles electronic structure and total energy calculations of the phase stability and dislocation properties of Pt$_{3}$Al to reveal the origins of its yield stress low temperature anomaly (LTA) \cite{Gornostyrev}. In addition, Yan et al. have researched the phase transition and formation enthalpies of Pt$_{3}$Al under high pressure. The results show that the cubic structure is stable compared to the tetragonal structure up to the pressure of 100~GPa and has excellent resistance to volume deformation under high pressure \cite{Liu}. Despite the above investigations, there have been no systemic experimental or theoretical reports on the elastic and thermodynamic properties of L1$_{2}$ phase Pt$_{3}$Al alloys under high pressure.

As we know, high pressure leads to the phase transition and changes the physical and chemical properties of a solid, such as mechanical and thermodynamic properties, which are essential for a profound understanding the application of Pt$_{3}$Al alloy \cite{Pickard, Khazaei}. The elastic constants determine the response of a crystal to external forces and provide important information on the brittleness, ductility, anisotropy, and the resistance to deformation \cite{Ning, Ivashchenko}. A comprehensive understanding of the bulk modulus $B$, shear modulus $G$, Young's modulus $Y$ and Poisson's ratio plays an important role in determining the mechanical properties of solid materials. Furthermore, to better understand the thermodynamic properties of Pt$_{3}$Al under high pressure, the quasi-harmonic Debye model was adopted. Then, we discuss the obtained thermodynamic parameters including Debye temperature $\Theta_\textrm{D}$, specific heats $C_{P}$, thermal expansion coefficient $\alpha$, and Gr\"{u}neisen parameter $\gamma$.

Therefore, it is highly desirable to understand the physical, mechanical, and thermal properties of L1$_{2}$ phase Pt$_{3}$Al. In this paper, we performed a systematic investigation of the structural, elastic, and thermodynamic properties on the Pt$_{3}$Al alloy with the L1$_{2}$ structure by first-principle calculations, and the calculated results were discussed in comparison with the available theoretical and experimental data.

\section{Methods}

In the present work, all the calculations were performed based on the plane wave pseudopotential density-function theory  method as implemented in CASTEP package \cite{Kresse}. The exchange correlation energy is described in the generalized gradient approximation (GGA) for the exchange correlation functional. Pt $5d^{9}4s^{1}$ and Al $3s^{2}3p^{1}$ were treated as valence electrons. A plane wave cutoff energy of 400~eV was employed. The Brillouin zone was sampled by a $14\times14\times14$ uniform $k$-point mesh according to the Monkhorst-Pack scheme grids.

In this work, the quasi-harmonic Debye model \cite{Blanco} implemented in the Gibbs program is used to obtain the thermodynamic properties of Pt$_{3}$Al. This model is sufficiently flexible in giving all thermodynamic quantities by incorporating the obtained results of energy and volume. The non-equilibrium Gibbs function $G^{*}(V;P,T)$ is described in the following form:
\begin{equation}\label{(1)}
   {G^*}(V;P,T) = E(V) + PV + {A_\textrm{vib}}(\Theta ,T).
\end{equation}
Here, $E(V)$ represents total energy/formula of Pt$_{3}$Al, $P$ is the hydrostatic pressure, $A_\textrm{vib}(\Theta,T)$ is used to represent lattice vibration Helmholtz free energy and is taken as:
\begin{equation}\label{(2)}
   {A_\textrm{vib}}(\Theta ,T) = nk_\text{B}T\left[ {\frac{{9\Theta }}{{8T}} + 3\ln \left(1 - {\re^{ - {\Theta }/{T}}}\right) - D({\Theta }/{T})} \right],
\end{equation}
where $D(\Theta/T)$ stands for the Debye integral, $n$ is the number of atoms per formula unit, and $\Theta$ is expressed by
\begin{equation}\label{(3)}
   \Theta  = \frac{\hbar }{k_\text{B}}{\left(6{\pi ^2}{V^{\frac{1}{2}}}n\right)^{\frac{1}{3}}}f(\nu)\sqrt {\frac{{{B_\textrm{S}}}}{M}}\,.
\end{equation}
In relation (\ref{(3)}), $M$ is the molecular mass per formula unit, $B_\textrm{S}$ is a representative for adiabatic bulk modulus, which is estimated in terms of static compressibility by using the following relation:
\begin{equation}\label{(4)}
   {B_\textrm{S}} = V\left(\frac{{{\rd^2}E(V)}}{{\rd{V^2}}}\right).
\end{equation}
And $f(\nu)$ is defined as follows:
\begin{equation}\label{(5)}
   f(\nu ) = {\left\{ {3{{\left[ {2{{\left(\frac{{21 + \nu }}{{31 - \nu }}\right)}^{3/2}} + {{\left(\frac{1}{3}\frac{{1 + \nu }}{{1 - \nu }}\right)}^{3/2}}} \right]}^{ - 1}}} \right\}^{1/3}},
\end{equation}
where $\nu$ is Poisson's ratio. Hence, the  non-equilibrium Gibbs function $G^{*}(V;P,T)$ as a function of $(V;P,T)$ can be minimized with respect to the volume as follows:
\begin{equation}\label{(6)}
   {\left( {\frac{{\rd{G^*}(V;P,T)}}{{\rd V}}} \right)_{P,T}} = 0.
\end{equation}
In order to obtain the thermal equation of states, we should solve the equation (\ref{(6)}). After the equilibrium state of a given $V(P,T)$ has been obtained, the isothermal bulk modulus and other thermodynamic properties, such as the heat capacity, vibrational internal energy, and thermal expansion can be evaluated using the relations given as below:
\begin{equation}\label{(7)}
   {C_V} = 3nk_\text{B}\left[ {4D({\Theta }/{T}) - \frac{{3\Theta /T}}{{{\re^{\Theta /T}} - 1}}} \right],
\end{equation}
\begin{equation}\label{(8)}
   {C_P} = 3nk_\text{B}\left[ {4D({\Theta }/{T}) - \frac{{3\Theta /T}}{{{\re^{\Theta /T}} - 1}}} \right](1 + \alpha \gamma T),
\end{equation}
\begin{equation}\label{(9)}
   \alpha  = \frac{{\gamma {C_V}}}{{BV}},
\end{equation}
\begin{equation}\label{(10)}
   \gamma  =  - \frac{{\rd\ln \Theta (V)}}{{\rd\ln V}}\,,
\end{equation}
where $\gamma$ is the Gr\"{u}neisen parameter. This method has already been successfully used to investigate the thermodynamic properties of a series of compounds.

\section{Results and discussions}
\subsection{Structure property }
As we know, Pt$_{3}$Al has two kinds of structures including cubic phase and tetragonal phase. For cubic phase, Pt$_{3}$Al alloy has a Cu$_{3}$Au-type structure (space group: Pm3m, No:~221), with lattice parameters: $a = b = c = 3.876$~{\AA} \cite{Zhang}. The Pt and Al atoms are located at the site (0, 0, 0) and (0.5, 0, 0), respectively. Each Al atom is surrounded by twelve Pt atoms. For tetragonal phase, Pt$_{3}$Al has a space group: P4/mmm (No:~123) with experimental lattice parameters: ${a} = \emph{b} = 3.832$~{\AA} and ${c} = 3.894$~{\AA}. There are two types of Pt: 1c (0.5, 0.5, 0) and 2e (0, 0.5, 0.5), respectively. The Al atom is in the site 1a (0, 0, 0). We have calculated the formation enthalpies of tetragonal and cubic phase as the pressure increasing from 0 GPa to 100~GPa. Our results show that the formation enthalpies of cubic structure are lower than that of tetragonal structure below 100~GPa. This means that cubic Pt$_{3}$Al is stable under high pressure which is consistent with the result of Liu \cite{Liu}. To obtain equilibrium structural parameters, the atom position and structure of Pt$_{3}$Al were optimized. At 0~GPa, the calculated lattice parameters of cubic phase ${a}$ is 3.86~{\AA}. We note a very good agreement between our results and experimental data. This offers the reliability and accuracy to our further investigation.

\subsection{Elastic property}
\begin{figure}[!t]
\begin{center}
\includegraphics[width=0.95\textwidth]{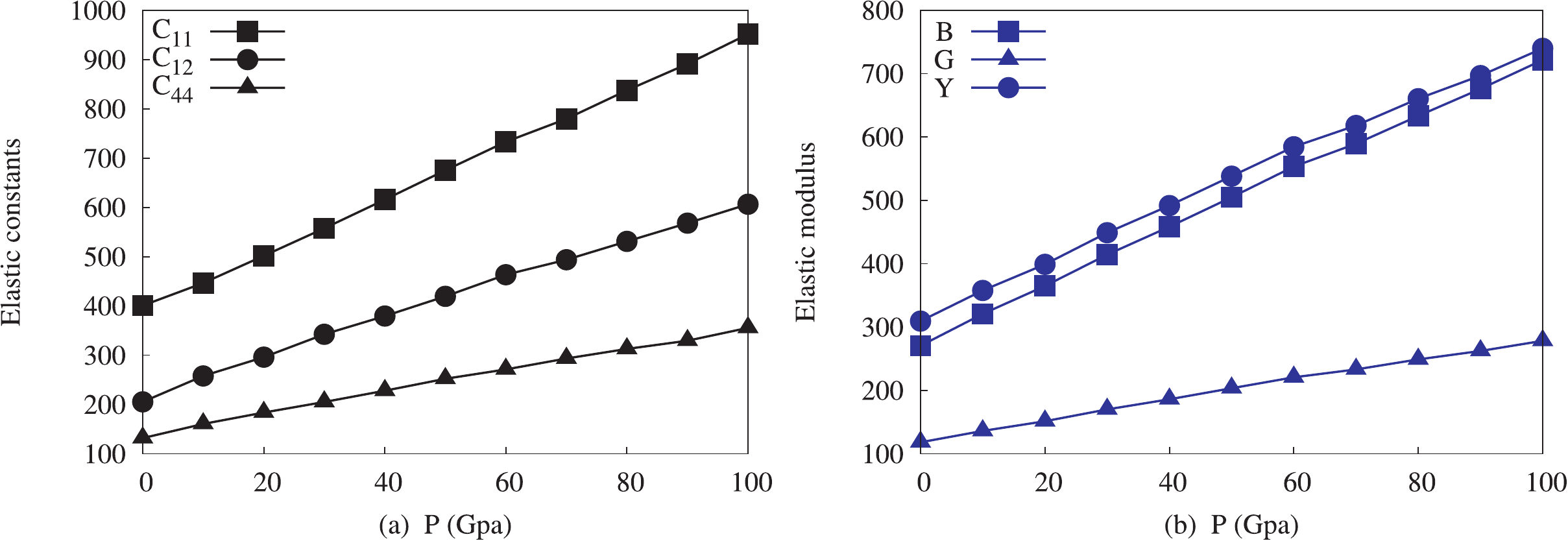}
\caption{The calculated elastic constants $C_{ij}$ and the elastic modulus of cubic Pt$_{3}$Al under pressure from 0~GPa to 100~GPa.}
\label{Fig1}
\end{center}
\end{figure}

To the best of our knowledge, the elastic properties define the behavior of a solid under different stress and strain conditions. The elastic stiffness parameters can describe the bonding characteristics, mechanical deformations, and structural stability \cite{Yang}. To obtain the elastic constants, a small strain should be loaded to the crystal. They can be got by calculating the total energy as a function of appropriate lattice deformation, which are expanded as the Taylor expansion for a system with respect to a small strain $\delta$ and volume $V_{0}$ \cite{Phillpot,Fast}. The elastic strain energy ${E(V)}$ is expressed as follows:
\begin{equation}\label{(11)}
   E(V) = E({V_0},0) + \frac{1}{2}\sum\limits_i^6 {\sum\limits_j^6 {{C_{ij}}{\delta _i}{\delta _j}} }\,.
\end{equation}
Here, $C_{ij}$ are elastic constants, $\delta_{i}$ and $\delta_{j}$ are related to the strain on the crystal. For cubic symmetry, there are three independent elastic constants, that are $C_{11}$, $C_{12}$, $C_{44}$. The calculated elastic constants $C_{ij}$ of Pt$_{3}$Al are shown in figure~\ref{Fig1}. At 0~GPa, the calculated elastic constants of Pt$_{3}$Al ($C_{11}=400.8$, $C_{12}=205.27$, $C_{44}=131.71$, and ${B}=270.46$) are consistent  with the experimental values ($B=277$) and other theoretical results ($C_{11}=395$, $C_{12}=210$, $C_{44}=118$)  \cite{Chauke}. In general, the requirements of mechanical stability in a cubic crystal lead to the following restrictions on the elastic constants: $C_{11}> 0$, $C_{12}> 0$, $C_{11}-C_{12}> 0$, $C_{11}+2C_{12}> 0$ \cite{Beckstein,Pan}. Obviously, our results in figure~\ref{Fig1}~(a) show that all the elastic constants satisfy the stabilities criteria up to 100~GPa. This clearly indicates that Pt$_{3}$Al under high pressure possesses mechanical stabilities. There is no doubt that the elastic constants of a solid are strongly affected by the pressure. It should be noted  that the elastic constants $C_{11}$, $C_{12}$, $C_{44}$ increase linearly with the pressure increase because the lattice parameters of Pt$_{3}$Al become shorter under pressure.

It is acknowledged that bulk modulus $B$ and shear modulus $G$ can measure the hardness in an indirect way. The calculated bulk modulus $B$, shear modulus $G$, and Young's modulus $Y$ under different pressures are shown in figure~\ref{Fig1}~(b). It is found that bulk modulus $B$, shear modulus $G$, and Young's modulus $Y$ of Pt$_{3}$Al  gradually increase as pressure increases, indicating that Pt$_{3}$Al becomes more and more difficult to be compressed as the pressure increases. In addition, all the elastic modulus can be used as a measure of the average bond strength of atoms for a given crystal. A larger bulk modulus $B$ and Young's modulus $Y$ respond to the more covalent and stronger bond strength. Hence, it can be expected that Pt and Al atom can form covalent bonds under high pressure. The shear modulus $G$ is the relationship between the resistance to reversible deformations and the shear stress. A high shear modulus $G$ is mainly due to the elastic constants $C_{44}$, because a large $C_{44}$ implies a stronger resistance to shear in the (1\,0\,0) plane.

\begin{table}[!b]
\caption{The Poisson's ratio $\nu$ and $B/G$ of Pt$_{3}$Al under high pressure.}
\begin{center}
\begin{tabular}  {c c c c c c c c c c c c}
\hline\hline Pressure &0&10&20&30&40&50&60&70&80&90&100\\
(GPa) &&&&&&&&&&\\
 \hline
$B/G$&2.289&2.356&2.410&2.436&2.462&2.480&2.506&2.527&2.543&2.577&2.592\\
$\nu$&0.309&0.314&0.318&0.319&0.321&0.322&0.324&0.325&0.326&0.328&0.329\\
\hline\hline
\end{tabular}
\end{center}
\label{Table1}
\end{table}

For a further analysis, the deformation behavior of Pt$_{3}$Al, the value of $B/G$ and Poisson's  ratio $\nu$ which are related with the brittleness and hardness of the materials are shown in table~\ref{Table1}. Generally, the $B/G$ ratio is used to predict the brittle or ductile behavior of materials. The critical value which separates ductile and brittle material is 1.75. The material exhibits  a ductile behaviour when the value $B/G > 1.75$; otherwise, the material behaves in a brittle manner \cite{Jafari, Kutepov}. We found that the $B/G$ is 2.89 at 0~GPa and increases with the pressure increase. It means that the Pt$_{3}$Al belongs to a ductility material and the pressure can improve the ductility of Pt$_{3}$Al. Another important property is the Poisson's ratio $\nu$ which is defined as the absolute value of the ratio of transverse strain to longitudinal strain. It is used to quantify the stability of the crystal against shear. The larger is the Poisson's ratio $\nu$, the better is the plasticity. This usually refers to ductile compounds with a large value ($>0.26$). It is noted that the Poisson's ratio $\nu$ increases from 0.309 to 0.329 with the pressure increase, indicating that Pt$_{3}$Al is ductile and the pressure can enhance the stability and the ductility of Pt$_{3}$Al. This result is consistent with that of the above elastic modulus ratio ($B/G$).

\subsection{Thermodynamic properties}

\begin{figure}[!h]
\begin{center}
\includegraphics[width=0.95\textwidth]{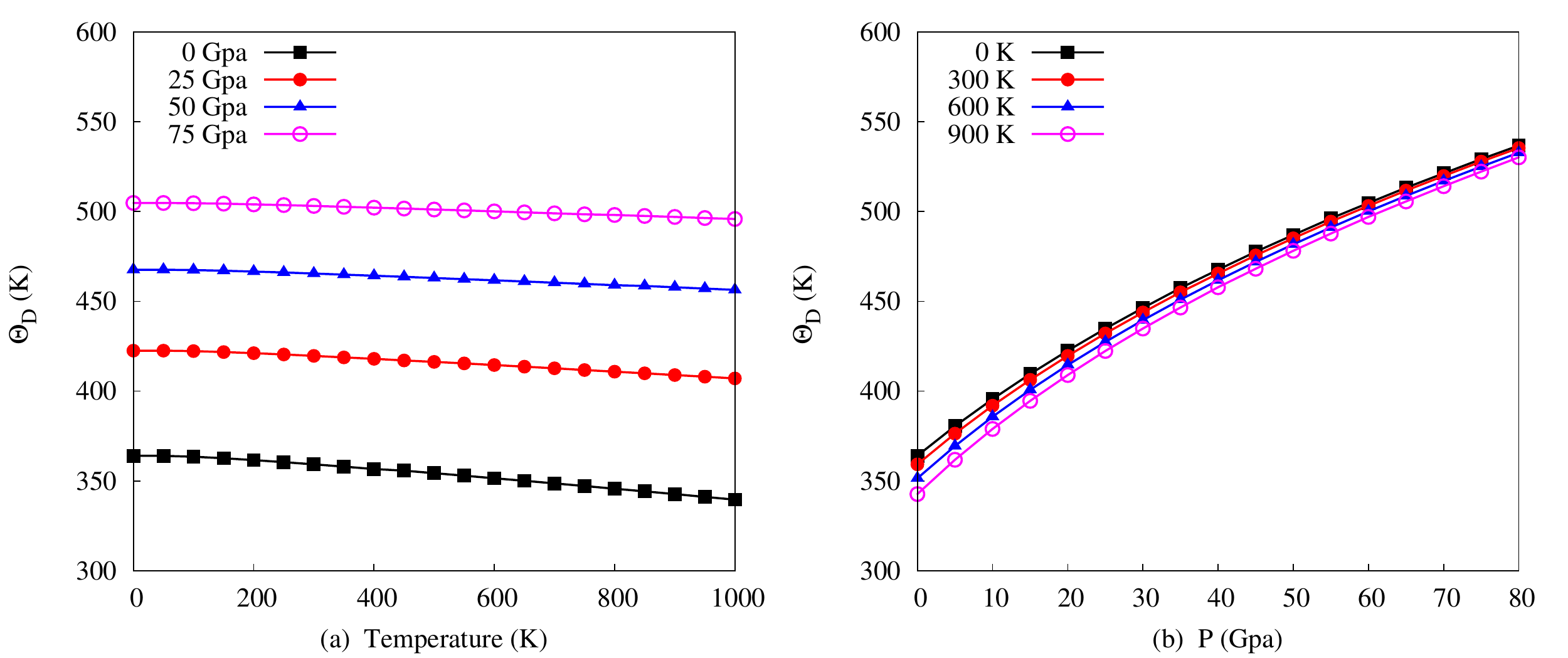}
\caption{(Color online) The Debye temperature as a function of temperature and pressure. (a) $P=0$, 25, 50, and 75~GPa, respectively; (b) ${T}=0$, 300, 600, and 900~K, respectively.}
\label{Fig2}
\end{center}
\end{figure}

To our knowledge, Debye temperature $\Theta_\textrm{D}$ is one of the most important parameters describing the thermal characteristics of compounds. The Debye temperature correlates with many physical properties of material, which are derived from elastic properties under pressures. Some detailed information of a solid, such as the melting temperature and specific heat can be found by calculating the Debye temperature. We obtain the thermodynamic properties of Pt$_{3}$Al at various temperatures and pressures from the energy-volume relations using the quasi-harmonic Debye model. The Debye temperature $\Theta_\textrm{D}$ as a function of temperature at different pressures is shown in figure~\ref{Fig2}~(a). It can be clearly seen that $\Theta_\textrm{D}$ in the range of temperatures from 0 to 1000~K approximately remains unaltered with the temperature increase, meaning that it is insensitive to temperature. Figure~\ref{Fig2}~(b) shows the Debye temperature as a function of pressure at different temperatures of $T=0$, 300, 600, and 900~K. It is noted that $\Theta_\textrm{D}$ linearly increases and further compression slows down the increase. As the pressure goes higher, the decreased magnitude of Debye temperature $\Theta_\textrm{D}$ becomes small. This is because the Debye temperature is related to the volume \emph{V} and adiabatic bulk modulus. Figure~\ref{Fig2}~(b) shows that when the temperature is constant, the Debye temperature $\Theta_\textrm{D}$ increases non-linearly with the applied pressures, indicating the change of the vibration frequency of particles under pressure. Hence, the temperature has a more significant effect on the Debye temperature $\Theta_\textrm{D}$ than pressure, and the temperature exhibits a smaller and smaller effect on the Debye temperature with an increase of pressure.

\begin{figure}[!t]
\begin{center}
\includegraphics[width=0.95\textwidth]{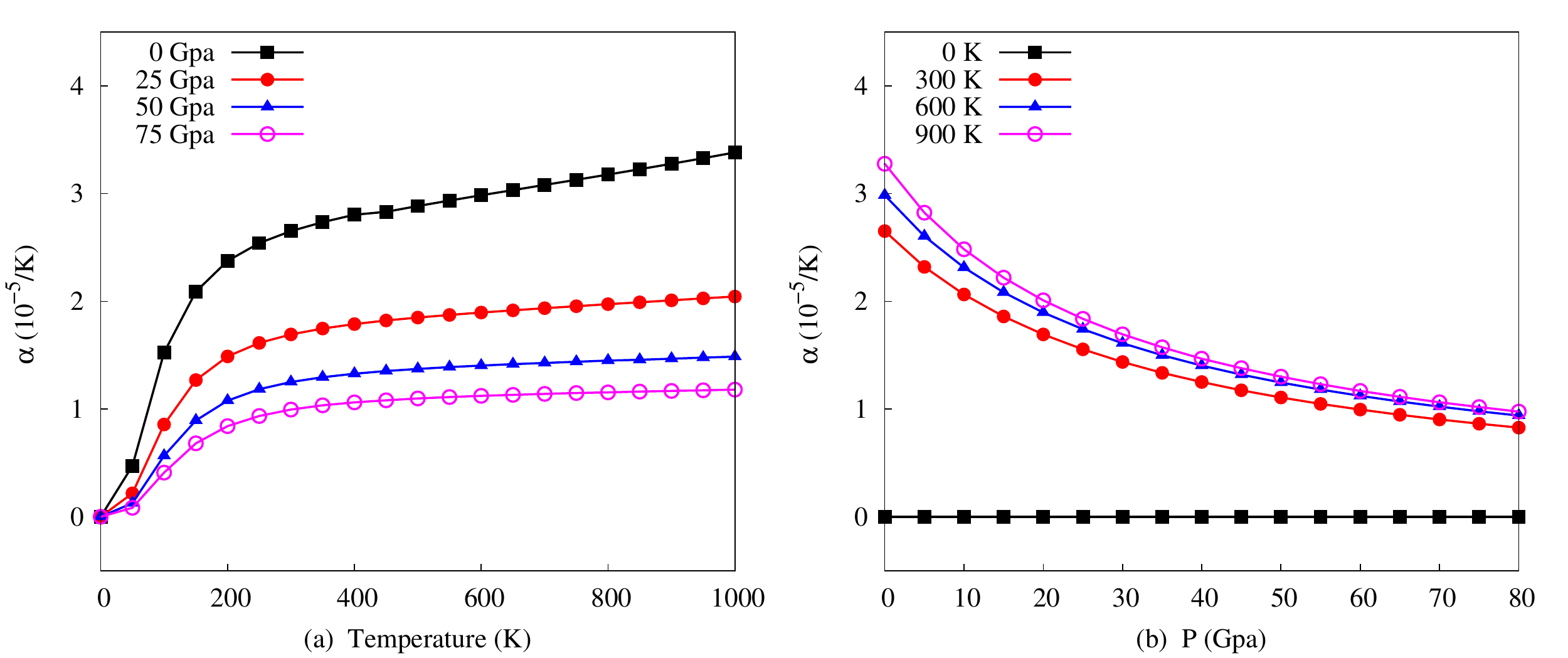}
\caption{(Color online) The volume thermal expansion coefficient as a function of temperature and pressure. (a) $P=0$, 25, 50, and 75~GPa, respectively;
(b) ${T}=0$, 300, 600, and 900~K, respectively.}
\label{Fig3}
\end{center}
\end{figure}

To describe the thermal properties of a material, the volume thermal expansion coefficient $\alpha$ is another essential parameter.  The dependence of the volume thermal expansion coefficient $\alpha$ of Pt$_{3}$Al on the temperature and pressure is illustrated in figure~\ref{Fig3}. We noted that $\alpha$ increases rapidly with $T^{3}$ at zero or low pressure when the temperature is below 200 K and gradually approaches a very low linear increase above 400~K for a given pressure. Moreover, we can also see that the values of $\alpha$ at zero pressure are much greater than those at other pressures. Figure~\ref{Fig3}~(b) gives $\alpha$ as a function of pressure at different temperatures of ${T}=0$, 300, 600, 900~K. It can be seen that for a given temperature, the thermal expansion coefficient $\alpha$ is zero at 0~K and rapidly decreases with the pressure increase, and it becomes flat under high pressure. Moreover, the higher is the temperature, the faster the $\alpha$ decreases. There is observed a larger thermal expansion at a higher temperature and at a lower pressure, and it provides less sensitivity of $\alpha$ at high temperature and high pressure for Pt$_{3}$Al.

\begin{figure}[!b]
\begin{center}
\includegraphics[width=0.95\textwidth]{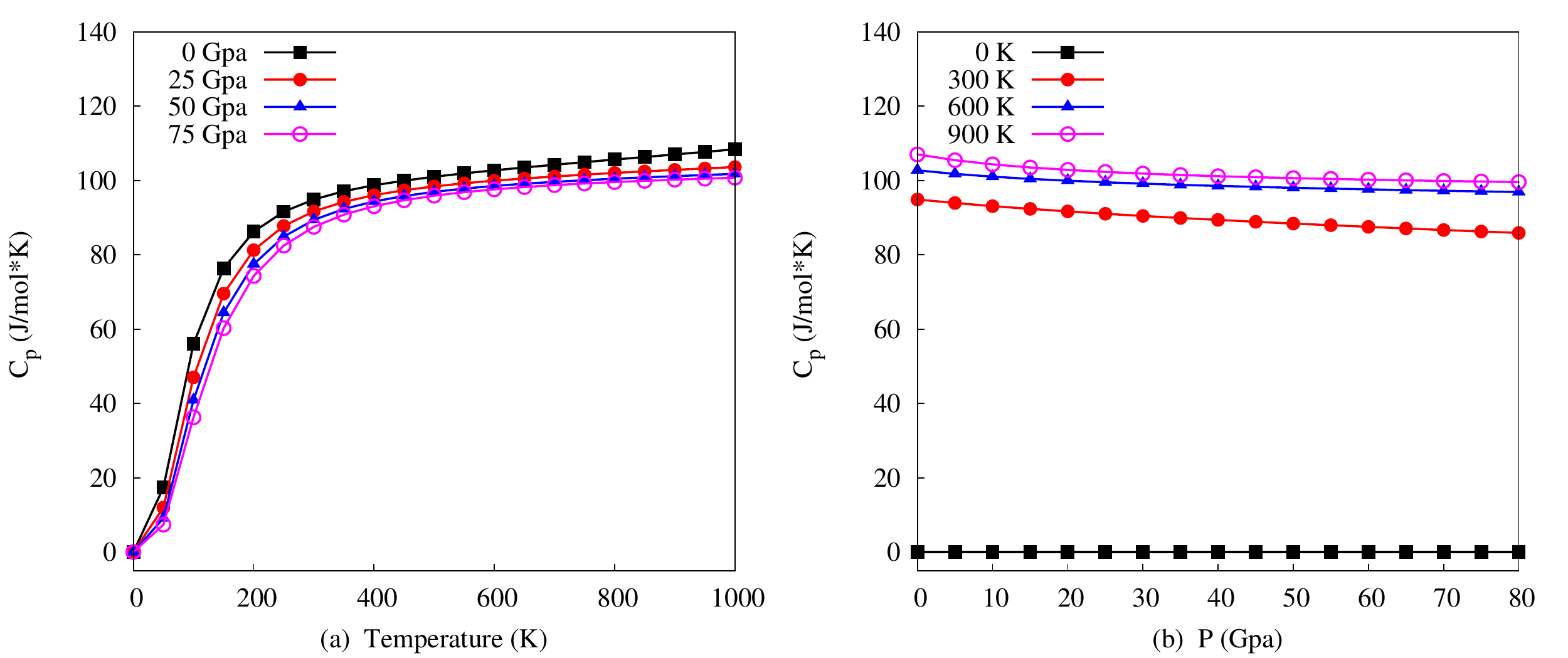}
\caption{(Color online) The heat capacity as a function of temperature and pressure. (a) $P=0$, 25, 50, and 75~GPa, respectively; (b) $T=0$, 300, 600, and 900~K, respectively.}
\label{Fig4}
\end{center}
\end{figure}

As another important thermodynamic parameter of solids, the heat capacity $C_{P}$ is of key importance for linking thermodynamics with microscopic structures and dynamics. Moreover, the knowledge of the heat capacity of a substance not only provides an essential insight into its vibrational properties but also is mandatory for many applications. Figure~\ref{Fig4} shows the calculated heat capacity $C_{P}$ as a function of temperature and pressure. It is obvious that $C_{P}$ follows the relationship of the Debye model [$C(T) \propto  T^{3}$] up to 200~K. Then, it monotonously increases with the temperature increase and converges to a constant Dulong-Petit limit, which is common to all solids at high temperatures. We note that the heat capacity $C_{P}$ slowly decreases  with the pressure increase, and the high temperature will slow down this trend seen in figure~\ref{Fig4}~(b). Figure~\ref{Fig4} implies that temperature and pressure have an opposite effect on the heat capacity, while the temperature has a greater effect on the heat capacity than the pressure.

\begin{figure}[!t]
\begin{center}
\includegraphics[width=0.95\textwidth]{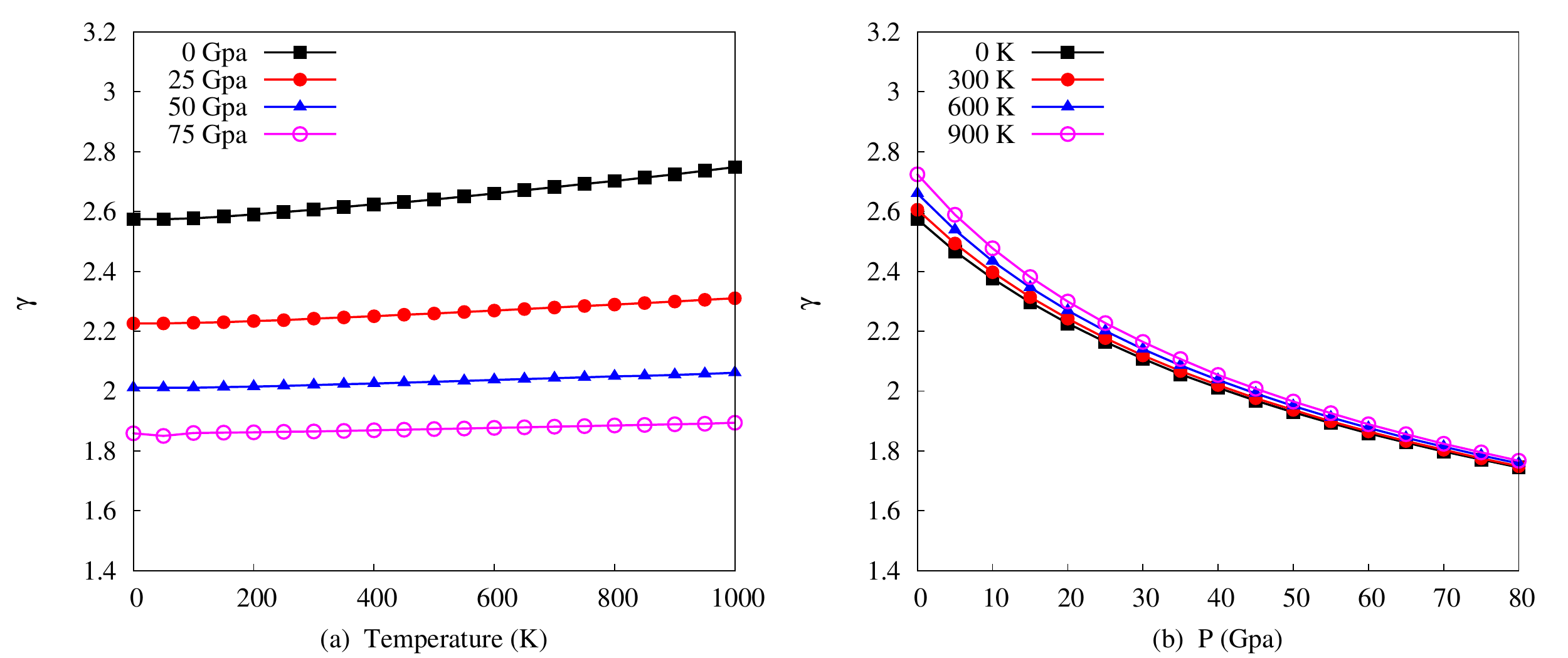}
\caption{(Color online) The Gr\"{u}neisen parameter $\gamma$ as a function of temperature and pressure. (a) $P=0$, 25, 50, and 75~GPa, respectively;
(b) $T=0$, 300, 600, and 900~K, respectively.}
\label{Fig5}
\end{center}
\end{figure}

In the quasi-harmonic Debye model, the Gr\"{u}neisen parameter $\gamma$ is of a great significance. It describes the anharmonic effects of the crystal lattice thermal vibration and has been widely used to characterize the thermodynamic behavior of a material at high pressure. The calculated Gr\"{u}neisen parameter $\gamma$ with pressure and temperature are presented in figure~\ref{Fig5}. It can be observed that the Gr\"{u}neisen parameter $\gamma$ almost keeps unchanged with the temperature increase at a fixed pressure in figure~\ref{Fig5}~(a), while it quickly decreases with the applied pressure. This is because the Gr\"{u}neisen parameter $\gamma$ is as function of the volume which is affected by the pressure in the quasi-harmonic model. And, there is a larger thermal expansion at low pressure. Those results suggest that the effect of the temperature on the Gr\"{u}neisen parameter $\gamma$ is not as significant as that of the pressure $P$. Furthermore, the Gr\"{u}neisen parameter $\gamma$ increases more slowly at high pressure than at low pressure.

\section{Conclusions}

First principles calculations are performed to investigate the elastic and thermodynamic properties of L1$_{2}$ phase Pt$_{3}$Al alloy under high pressure and high temperature. The elastic constants, bulk modulus $B$, shear modulus $G$, and Young's modulus $Y$ as a function of the pressure have been systematically investigated. The results show that all the elastic constants meet the corresponding mechanical stability criteria and the elastic modulus increases linearly with the applied pressure. The Poisson's ratio $\nu$ and the elastic modulus ratio ($B/G$) show that L1$_{2}$ phase Pt$_{3}$Al alloy  is found to be ductile and higher pressure can significantly enhance the ductility. This means that the elastic properties of Pt$_{3}$Al will be improved under high pressure. To study the thermal and vibrational effects, the quasi-harmonic Debye is used. The dependence of Debye temperature $\Theta_\textrm{D}$, specific heats $C_{P}$, thermal expansion coefficient $\alpha$, and the Gr\"{u}neisen parameter $\gamma$ are systematically explored in the ranges of 0--100~GPa and 0--1000~K. We find that the temperature has a more significant effect on the Debye temperature $\Theta_\textrm{D}$ and the heat capacity $C_{P}$ than pressure. Furthermore, the thermal expansion coefficient $\alpha$ becomes insensitive to high temperature and high pressure.

\section*{Acknowledgements}
This work was supported by the special Funds for Major State Basic Research Project of China (973) under grant No.~2012CB933702, the National Science Foundation of China under Grants Nos. NSAF \linebreak U1230202 and 11204310. Part of the calculations were performed in Center for Computational Science of CASHIPS, the ScGrid of Supercomputing Center, and Computer Network Information Center of Chinese Academy of Sciences.

\ukrainianpart

\title{Теоретичне дослідження пружних і термодинамічних властивостей Pt$_{3}$Al зі структурою L1$_{2}$  під впливом тиску}
\author{Н. Вей\refaddr{label1},
       Ч. Жанг\refaddr{label2}, С. Гоу\refaddr{label2}}
\addresses{
\addr{label1} Університет провінції Аньхой, м. Хефей, Китай
\addr{label2} Головна лабораторія матеріалознавства, Інститут фізики твердого тіла, Китайська академія наук, м.~Хефей, Китай
}

\makeukrtitle
\begin{abstract}
В цій роботі досліджуються пружні і термодинамічні властивості  Pt$_{3}$Al при високому тиску, використовуючи теорію функціоналу густини
з узагальненим градієнтним наближенням.
Результати для об'ємного модуля пружності і пружних констант  при нульовому тиску добре узгоджуються з наявними теоретичними і
експериментальними даними. При високому тиску всі пружні константи задовільняють відповідний критерій
механічної стійкості, що свідчить про механічну стійкість   Pt$_{3}$Al. Окрім цього, пружні константи і модуль пружності зростають лінійно
з прикладеним тиском.  У відповідності з коефіцієнтом Пуассона $\nu$ і коефіцієнтом модуля пружності ($B/G$), знайдено, що
сплав Pt$_{3}$Al є пластичним, і вищі тиски можуть значно посилити його пластичність. Це вказує на те, що пружні властивості Pt$_{3}$Al
будуть покращені при високому тиску.
В результаті використання квазігармонічної моделі Дебая, вперше отримано зміну температури
$\Theta_\textrm{D}$,
питомої теплоємності $C_{P}$, коефіцієнта теплового розширення $\alpha$ і параметра Грюнайзена $\gamma$ в області  тисків від
0 до 100~GPa і температур від 0 до 1000~K.
\keywords перші принципи, пружні властивості, термодинамічні властивості

\end{abstract}


\clearpage

\begin{thebibliography}{99}

\bibitem{Jiang}
Jiang C.,  Sordelet D.J., Phys. Rev. B, 2005, \textbf{72}, 184203; \bibdoi{10.1103/PhysRevB.72.184203}.

\bibitem{Yamabe}
Yamabe Y.,  Koizumi Y.,  Murakami H.,  Maruko Y.R.T.,  Harada H., Scr. Mater., 1996, \textbf{35}, 211; \\ \bibdoi{10.1016/1359-6462(96)00109-1}.

\bibitem{Meschel}
Meschel S.V.,  Nash P.,  Chen X.Q., J. Alloys Compd., 2010, \textbf{492}, 105; \bibdoi{10.1016/j.jallcom.2009.11.092}.

\bibitem{Wang}
Wang H.Y., Cao J., Condens. Matter Phys., 2012, \textbf{15}, 13705; \bibdoi{10.5488/CMP.15.13705}.

\bibitem{Wu}
Wu K., Jin Z., J. Phase Equilib., 2000, \textbf{21}, 221; \bibdoi{10.1361/105497100770340011}.

\bibitem{Feng}
Feng J.,  Xiao B.,  Chen J., Mater. Des., 2011, \textbf{32}, 3231;
\bibdoi{10.1016/j.matdes.2011.02.043}.

\bibitem{Duan}
Duan Y.H., Huang B., Sun Y., J. Alloys Compd., 2014, \textbf{590}, 50; \bibdoi{10.1016/j.jallcom.2013.12.079}.

\bibitem{Chauke}
Chauke H.R.,  Minisini B., Intermetallics, 2010, \textbf{18}, 417; \bibdoi{10.1016/j.intermet.2009.08.016}.

\bibitem{Norihiko}
Okamoto N.L., Hasegawa Y., Hashimoto W., Philos. Mag., 2013, \textbf{93}, 60; \bibdoi{10.1080/14786435.2012.705037}.

\bibitem{Gornostyrev}
Gornostyrev Yu.N.,  Kontsevoi O.Yu.,  Maksyutov A.F., Phys. Rev. B, 2004, \textbf{70}, 014102; \\ \bibdoi{10.1103/PhysRevB.70.014102}.

\bibitem{Liu}
Liu Y., Huang H., J. Alloys Compd., 2014, \textbf{597}, 200;
\bibdoi{10.1016/j.jallcom.2014.02.001}.

\bibitem{Pickard}
Pickard C.J.,  Needs R.J., Phys. Rev. Lett., 2006, \textbf{97}, 045504; \bibdoi{10.1103/PhysRevLett.97.045504}.

\bibitem{Khazaei}
Khazaei M.,  Tripathi M.N., Phys. Rev. B, 2011, \textbf{83}, 134111; \bibdoi{10.1103/PhysRevB.83.134111}.

\bibitem{Ning}
Wei N., Jia T., Zeng Z., AIP Adv., 2014, \textbf{4}, 057103; \bibdoi{10.1063/1.4875024}.

\bibitem{Ivashchenko}
Ivashchenko V.I.,  Turchi P.E.A.,  Shevchenko V.I., Condens. Matter Phys., 2013, \textbf{16}, 33602; \\ \bibdoi{10.5488/CMP.16.33602}.

\bibitem{Kresse}
Kresse G.,  Joubert D., Phys. Rev. B, 1999, \textbf{59}, 1758; \bibdoi{10.1103/PhysRevB.59.1758}.

\bibitem{Blanco}
Blanco M.A.,  Francisco E., Comput. Phys. Commun., 2004, \textbf{158}, 57; \bibdoi{10.1016/j.comphy.2003.12.001}.

\bibitem{Zhang}
Zhang Q.A.,  Akiba E., J. Alloys Compd., 2003, \textbf{360}, 143; \bibdoi{10.1016/S0925-8388(03)00320-7}.

\bibitem{Yang}
Yang Y., Lu H., Yu C., Chen J.M., J. Alloys Compd., 2009, \textbf{485}, 542; \bibdoi{10.1016/j.jallcom.2009.06.023}.

\bibitem{Phillpot}
Wang J.,  Li J.,  Yip S.,  Phillpot S., Phys. Rev. B, 1995, \textbf{52}, 12627; \bibdoi{10.1103/PhysRevB.52.12627}.

\bibitem{Fast}
Fast L.,  Wills J.M.,  Johansson B., Phys. Rev. B, 1995, \textbf{51}, 17431; \bibdoi{10.1103/PhysRevB.51.17431}.

\bibitem{Beckstein}
Beckstein O.,  Klepeis J.E.,  Hart G.L.W., Phys. Rev. B, 2001, \textbf{63},
134112; \bibdoi{10.1103/PhysRevB.63.134112}.

\bibitem{Pan}
Pan Y.,  Guan W.M.,  Zhang K.H., Physica B, 2013, \textbf{427}, 17; \bibdoi{10.1016/j.physb.2013.05.039}.

\bibitem{Jafari}
Jafari M.,  Nobakhti M.,  Jamnezhad H.,  Bayati K., Condens. Matter Phys., 2013, \textbf{16}, 33703; \\ \bibdoi{10.5488/CMP.16.33703}.

\bibitem{Kutepov}
Kutepov A.L., Kutepova S.G., Phys. Rev. B, 2003, \textbf{67}, 132102; \bibdoi{10.1103/PhysRevB.67.132102}.

\end{thebibliography}
\end{document}